\documentclass{article}
\usepackage[english]{babel}
\usepackage{amsmath}  
\usepackage{physics}   
\usepackage{authblk}
\usepackage{comment}
\usepackage{caption}
\usepackage{bm}
\usepackage[numbers,sort&compress]{natbib} 
\setcitestyle{square, comma, compress}  
\let\cite=\citep  
\date{}

\def\beq{\begin{equation}}
\def\eeq{\end{equation}}
\def\tb {t_\beta}
\def\sb  {s_{\beta}}
\def\cb  {c_{\beta}}

\usepackage[letterpaper,top=2cm,bottom=2cm,left=3cm,right=3cm,marginparwidth=1.75cm]{geometry}
\usepackage{graphicx}
\usepackage[pdfencoding=auto,unicode=true,colorlinks=true, allcolors=blue]{hyperref}

\title{Signature for charged Higgs pair production in 2HDM+a model at CLIC}
\author[1,2]{Shuo Yang\thanks{Email: shuoyang@lnnu.edu.cn}}
\author[1,2]{Ziyang Yu\thanks{Email: 18242541728@163.com}}
\author[1,2]{Yiqi Wang\thanks{Email: wyqwyq000@163.com}}
\author[3]{Lei Wang\thanks{Email: leiwang@yantai.edu.cn}}

\affil[1]{School of Physics and Electronic Technology, Liaoning Normal University, Dalian, 116029, China}
\affil[2]{Center for Theoretical and Experimental High Energy Physics, Liaoning Normal University, Dalian, 116029, China} 
\affil[3]{Physics Department, Yantai University, Yantai, 264005, China}

\begin{document}
\maketitle

\begin{abstract}
In this work, we study the search for charged Higgs bosons in the Two-Higgs-Doublet Model plus an additional pseudo-scalar (2HDM+a) at the Compact Linear Collider (CLIC).  Focusing on the pair production of charged Higgs bosons, followed by the decays \( H^\pm \to a W \) and \( H^\mp \to t b \), we analyze the signal channel of $4j+2b+E_T^{miss}$.
Given the center-of-mass energy of CLIC as \( \sqrt{s} = 1500 \, \text{GeV} \) , charged Higgs bosons with masses \( m_{H^\pm} < \sqrt{s}/2 \) are explored. 
Simulation results indicate that the signal significance of 400 GeV and 600 GeV charged Higgs bosons can reach $5\sigma$ in specific parameter spaces, and the $2\sigma$ exclusion limits in the $m_{H^\pm} - m_a$ plane are presented for $m_{H^\pm}$ in the range of [400, 650] GeV.
\end{abstract}

\section{Introduction}

The existence of dark matter (DM), while robustly supported by a wealth of astrophysical and cosmological measurements, remains poorly understood (For a recent review, see Ref~\cite{Cirelli:2024ssz}). The Standard Model (SM) of particle physics lacks the necessary ingredients to account for observed phenomena, necessitating the inclusion of viable DM candidates in many extensions beyond the Standard Model (BSM)~\cite{Cirelli:2024ssz, ParticleDataGroup:2024cfk, Buchmueller:2017qhf, Kahlhoefer:2017dnp, Boveia:2018yeb, Argyropoulos:2021sav, Arcadi:2024ukq, Avila:2025qsc}. One approach to the theory of DM is building interactions between DM particles and SM particles with an intermediate particle known as a mediator~\cite{Abdallah:2015ter, Abercrombie:2015wmb, No:2015xqa, Goncalves:2016iyg, Bauer:2017ota, Argyropoulos:2024yxo}. In many simplified models, the most commonly adopted assumption is the existence of a cosmologically stable DM particle and a mediator particle, where the mediator can be a vector, axial-vector, scalar, or pseudoscalar particle depending on the models~\cite{Abdallah:2015ter, Abercrombie:2015wmb, No:2015xqa, Goncalves:2016iyg, Bauer:2017ota, Argyropoulos:2024yxo, Boveia:2016mrp, Albert:2017onk}. These models are characterized by a minimal number of free parameters, usually the masses and coupling constants of the DM particle and the mediator particle.

Colliders, as the most powerful experimental tools in particle physics, have significantly deepen our understanding of ordinary matter and also bear the responsibility of unraveling the mysteries of DM.
In collider based DM searches, high-energy collisions of SM particles offer a unique avenue: they can not only produce DM directly under controlled experimental conditions but also grant access to mediator particles~\cite{Buchmueller:2017qhf, Kahlhoefer:2017dnp, Boveia:2018yeb, Argyropoulos:2021sav, Arcadi:2024ukq, Avila:2025qsc}. The large hadron collider (LHC) experiments, especially ATLAS and CMS groups, explore DM production through mediator particles, which manifest as missing transverse energy (MET) signatures accompanied by visible SM particles~\cite{Buchmueller:2017qhf, Kahlhoefer:2017dnp, Abdallah:2015ter, Abercrombie:2015wmb, No:2015xqa, Goncalves:2016iyg, Bauer:2017ota, Argyropoulos:2024yxo, Boveia:2016mrp, Albert:2017onk}. 

On the other hand, following the discovery of the Higgs boson, the exploration of the scalar sector of the SM and its extensions within new physics frameworks have emerged as a primary focus for the LHC and future colliders. A particularly natural framework that accommodates a DM candidate while addressing some key open issues of the SM, such as the electroweak hierarchy problem~\cite{Golfand:1971iw, Volkov:1973ix, Wess:1974tw, Wess:1974jb, Ferrara:1974pu, Salam:1974ig}, baryogenesis~\cite{MCLERRAN1991477, Turok:1990zg, Cohen:1991iu, Cline:1996mga, Fromme:2006cm, Cline:2011mm} or the strong CP problem ~\cite{Kim:1986ax}, is the Two-Higgs-Doublet Model augmented with a pseudoscalar, denoted as 2HDM+a. The 2HDM+a forms a simple, ultraviolet-complete, gauge-invariant, and renormalizable extension of the simplified pseudoscalar mediator framework~\cite{Abercrombie:2015wmb, Argyropoulos:2024yxo, Buckley:2014fba}. The model is highly suitable for both direct and indirect searches. This is attributed to the pseudoscalar mediator being subject to relatively mild constraints from direct detection experiments, while simultaneously reproducing the observed relic abundance over a large region of the model parameter space~\cite{Bauer:2017ota, Argyropoulos:2024yxo}.

Furthermore, the 2HDM+a predicts rich collider phenomena~\cite{Bauer:2017ota, Argyropoulos:2024yxo, LHCDarkMatterWorkingGroup:2018ufk, Robens:2021lov, Argyropoulos:2022ezr, Dutta:2025nmy, Darvishi:2025gdl}, even including some features not foreseen in commonly used simplified models. It has now been recognized by the LHC Dark Matter Working Group as a benchmark model~\cite{Kahlhoefer:2017dnp, LHCDarkMatterWorkingGroup:2018ufk}. In addition to the fermionic DM candidate $\chi$, the 2HDM+$a$ model postulates the existence of five additional Higgs particles: a CP-even scalar $H$, two pseudoscalars $A$ and $a$, and a pair of charged Higgs bosons $H^\pm$~\cite{Bauer:2017ota, Argyropoulos:2024yxo, Robens:2021lov}. 
Also, the 2HDM+$a$ model can realize a variety of electroweak phase transition scenarios \cite{Liu:2023sey, Si:2024vrq}.
The charged Higgs boson is a charming particle drawn considerable attention, whose signature is the direct evidence of two-Higgs-doublet scalar structure \cite{Branco:2011iw, Yang:2011jk, Guchait:2018nkp, Li:2024kpd, Hashemi:2024zvg, Hashemi:2023osd, Ouazghour:2024twx, Duarte:2024zeh, Yue:2024bsk, Coleppa:2025hmf, Sun:2025wxn}. Furthermore, the introduction of the pseudoscalar in 2HDM+a bring a new decay channel $H^+\rightarrow aW^+$, which induce distinguished signature at colliders \cite{Bauer:2017ota, Argyropoulos:2024yxo}.

Currently, there are two main specific incarnations of 2HDM+a according to the Yukawa interaction structure of two higgs doublets. In Type II 2HDM+a \cite{Bauer:2017ota}, the Yukawa sector is type II, which has been extensively studied. Generally, the constraints arising from direct searches for BSM Higgs bosons are weaker in type-I than in type-II 2HDM models.
This is because the couplings of $H$, $A$, $a$, $H^{\pm}$ to charged leptons, down-type quarks and up-type quarks are suppressed by a factor of $1/\tan\beta$ in alignment limit in Type I structure. As discussed in Ref \cite{Argyropoulos:2024yxo}, 2HDM+a model with a type I Yukawa sector can accommodate a large region of parameter space, giving rise to experimental signatures that remain completely unexplored. In this paper, we take Type I 2HDM+a as the show case and study the pair production of the charged Higgs in alignment limit and small $\tan\beta$ region at the future Compact Linear Collider (CLIC) with center-of-mass energy of 1.5 TeV. Actually, in small $\tan\beta$ region, the cross sections and decay ratios for charged Higgs are nearly the same for both types of 2HDM+a models\cite{Bauer:2017ota, Argyropoulos:2024yxo}. 
We consider one charged Higgs boson decaying as $H^{\pm}\rightarrow tb$ and the other as $H^{\mp}\rightarrow aW^\mp$ followed by the decays $t\to jjb$, $W\to jj$, and $a\to \chi \chi$, resulting in
the signature $4j+2b+E_T^{miss}$. The clean collision environment at CLIC, combined with rapid advancements in particle detection techniques, makes the identification of this charged Higgs signal feasible at this planned multi-TeV energy frontier collider \cite{Adli:2025swq}. 
The structure of this paper is organized as follows: In Section 2, we present a brief introduction of the theoretical framework of the 2HDM+$a$ model. In Section 3, we calculate the branching ratios of charged Higgs. Furthermore, we carry out a full simulation and analyze the singal channel $e^+e^-\to H^+H^- \to tbaW \to 4j+2b+E_T^{miss}$. Finally, conclusions and summaries are provided in Section 4.

\section{Brief Introduction of 2HDM + a}

In this section, we give a brief overview of the 2HDM+a model \cite{Bauer:2017ota, Argyropoulos:2024yxo, Robens:2021lov, XENON:2019rxp}.
The full scalar potential consists of the standard 2HDM part supplemented by additional pseudoscalar singlet interactions:
\begin{equation}
V_{2HDM+a} = V_{2HDM} + V_{HP}.
\label{eq:1}
\end{equation}
We denote the 2HDM sector of the scalar potential as $V_{\text{2HDM}}$, given by \cite{Argyropoulos:2024yxo}

\begin{equation}
\begin{aligned}
V_{2HDM} = &\mu_{1}H_{1}^{\dagger}H_{1} + \mu_{2}H_{2}^{\dagger}H_{2} - \left(\mu_{3}H_{1}^{\dagger}H_{2} + \mathrm{h.c.}\right) + \frac{\lambda_{1}}{2}\left(H_{1}^{\dagger}H_{1}\right)^{2} + \frac{\lambda_{2}}{2}\left(H_{2}^{\dagger}H_{2}\right)^{2} \\
&+ \lambda_{3}\left(H_{1}^{\dagger}H_{1}\right)\left(H_{2}^{\dagger}H_{2}\right) + \lambda_{4}\left(H_{1}^{\dagger}H_{2}\right)\left(H_{2}^{\dagger}H_{1}\right) + \left[\lambda_{5}\left(H_{1}^{\dagger}H_{2}\right)^{2} + \mathrm{h.c.}\right].
\end{aligned}
\label{eq:2}
\end{equation}

$H_{1}$ and $H_{2}$ represent the two Higgs doublets,
\begin{equation}
H_1=\left(\begin{array}{c} \phi_1^+ \\
\frac{1}{\sqrt{2}}\,(v_1+\phi_1+i\eta_1)
\end{array}\right)\,, \ \ \
H_2=\left(\begin{array}{c} \phi_2^+ \\
\frac{1}{\sqrt{2}}\,(v_2+\phi_2+i\eta_2)
\end{array}\right).
\end{equation}
The vacuum expectation values (VEVs) of the two doublets are denoted by
 $v_1$ and $v_2$, with $v = \sqrt{v_1^2 + v_2^2} \approx 246\,\text{GeV}$, and their ratio defined as $\tan\beta = \frac{v_2}{v_1}$. To prevent tree-level flavor-changing neutral currents, a $Z_2$ symmetry is introduced, under which the transformations $H_1 \to H_1$ and $H_2 \to -H_2$ hold. With the exception of the soft-breaking term  $\mu_{3}H_{1}^{\dagger}H_{2} + \mathrm{h.c.}$, all interactions in $V_{H}$ preserve this symmetry \cite{Guchait:2018nkp}.

$V_{HP}$ represents the contribution to the potential from the singlet pseudoscalar,
\begin{equation}
    V_{HP} = \frac{1}{2} m_{P}^2 P^2 + \frac{1}{24} \kappa_{P} P^4+ P\left(i b_{P} H_1^\dagger H_2 + \text{h.c.}\right) + P^2 \left(\frac{\lambda_{P1}}{2} H_1^\dagger H_1 + \frac{\lambda_{P2}}{2} H_2^\dagger H_2\right).
    \label{eq:4}
\end{equation}
The singlet pseudoscalar $P$ has no VEV and transforms as
$P \to P$ under the $Z_2$ symmetry. The third term in $V_{HP}$ constitutes a soft breaking of this symmetry. The quartic term $P^4$ in $V_{HP}$ is a deliberate design, which has negligible influence on the phenomenology at the LHC~\cite{Goncalves:2016iyg, Ipek:2014gua}.

To avoid potential issues with electric dipole moments, all parameters in $V_{2HDM}$ and $V_{HP}$ are taken to be real, ensuring CP conservation in $V_{HP}$.
The minimization conditions of the potential require
\beq
\begin{split}
&\quad \mu_{1} = \mu_{3} \tb - \frac{1}{2} v^2 \left( \lambda_1 \cb^2 + \lambda_{345}\sb^2 \right)\,,\\
& \quad \mu_{2} =  \mu_{3} / \tb - \frac{1}{2} v^2 \left( \lambda_2 \sb^2 + \lambda_{345}\cb^2 \right)\,,
\end{split}
\label{min_cond}
\eeq
where the abbreviation $t_\beta\equiv \tan\beta$, $s_\beta\equiv \sin\beta$,  $c_\beta \equiv \cos\beta$, and $\lambda_{345} = \lambda_3+\lambda_4+\lambda_5$.

 After spontaneous symmetry breaking, the physical Higgs spectrum is obtained by diagonalizing the mass matrices. It consists of two CP-even states ($h$, $H$), two CP-odd states ($A$ and $a$), and a pair of charged Higgs bosons ($H^\pm$).
 The scalar masses can be chosen as input parameters to determine other relevant parameters,
\begin{eqnarray}\label{poten-cba}
&& b_p = \frac{m_{a}^2-m_A^2}{v}s_\theta c_\theta\,,\nonumber\\
&& m_P^2 = m_A^2 s_\theta^2+ m_a^2 c_\theta^2  -\frac{\lambda_{P1}}{2} v^2 \cb^2 - \frac{\lambda_{P2}}{2} v^2 \sb^2\,.\nonumber\\
 &&v^2 \lambda_1  = \frac{m_H^2 c_\alpha^2 + m_h^2 s_\alpha^2 - \mu_{3} t_\beta}{ c_\beta^2}, \ \ \ 
v^2 \lambda_2 = \frac{m_H^2 s_\alpha^2 + m_h^2 c_\alpha^2 - \mu_{3} t_\beta^{-1}}{s_\beta^2},  \nonumber \\  
&&v^2 \lambda_3 =  \frac{(m_H^2-m_h^2) s_\alpha c_\alpha + 2 m_{H^{\pm}}^2 s_\beta c_\beta - \mu_{3}}{ s_\beta c_\beta }, \ \ \ 
v^2 \lambda_4 = \frac{(\hat{m}_A^2-2m_{H^{\pm}}^2) s_\beta c_\beta + \mu_{3}}{ s_\beta c_\beta },  \nonumber \\
 &&v^2 \lambda_5=  \frac{ - \hat{m}_A^2 s_\beta c_\beta  + \mu_{3}}{ s_\beta c_\beta }\, , 
 \label{eq:lambdas}
\end{eqnarray}
with $\hat{m}_A^2=m_A^2 c_\theta^2+m_a^2 s_\theta^2$. The mixing angle $\alpha$ characterizes the CP-even states $h$ and $H$, while $\theta$ describes the mixing between $A$ and $a$. The shorthand notations $s_\alpha\equiv \sin\alpha$, $c_\alpha \equiv \cos\alpha$, $s_\theta\equiv \sin\theta$, $c_\theta \equiv \cos\theta$.

The interactions between a gauge boson and two scalar fields are described by the Lagrangian term,
\begin{align}
\mathcal{L}_{SSV} = &\frac{g}{2}W^+_\mu\left(c_\theta(H^-\overset{\leftrightarrow}{\partial}^\mu A)+s_\theta(H^-\overset{\leftrightarrow}{\partial}^\mu a)\right.\notag\\
&\left.-i\cos(\beta-\alpha)(H^-\overset{\leftrightarrow}{\partial}^\mu h)+i\sin(\beta-\alpha)(H^-\overset{\leftrightarrow}{\partial}^\mu H)+h.c. \right)\notag\\
&+\frac{g}{2c_W}Z_\mu\left(\cos(\beta-\alpha)(c_\theta A\overset{\leftrightarrow}{\partial}^\mu h + s_\theta a\overset{\leftrightarrow}{\partial}^\mu h) \right.\notag\\
&\left.-\sin(\beta-\alpha)(c_\theta A\overset{\leftrightarrow}{\partial}^\mu H + s_\theta a\overset{\leftrightarrow}{\partial}^\mu H)\right)\notag\\
&+\left(ie\gamma_\mu+i\frac{g(c_W^2-s_W^2)}{2c_W}Z_\mu \right)(H^\mp\overset{\leftrightarrow}{\partial}^\mu H^\pm),
\end{align}
where $c_W = \cos\theta_W$ and $s_W = \sin\theta_W$ with $\theta_W$ denoting the Weinberg angle.

For a Dirac DM field $\chi$ with mass $m_{\chi}$ that transforms as $\chi \to -\chi$ under the $Z_2$ symmetry, the symmetry allows only the following renormalizable coupling to the mediator,
\begin{equation}
    \mathcal{L}_{\chi} = -i y_{\chi} P \bar{\chi} \gamma_{5} \chi.
    \label{eq:8}
\end{equation}
Here the Yukawa coupling $y_{\chi}$ is taken to be real, and the interaction term $\mathcal{L}_{\chi}$ respects the $Z_2$ symmetry under which $\chi \to -\chi$. All BSM interactions in the 2HDM + a model are CP-conserving.

In the 2HDM + a model, in addition to the parameters of SM, the model introduces 14 parameters
\cite{Argyropoulos:2024yxo}:

\begin{equation}
\begin{array}{c}
\mu_{1},\mu_{2},\mu_{3}, b_{P},m_{P},m_{\chi},\\
y_{\chi},\lambda_{1},\lambda_{2},\lambda_{3},\lambda_{4},\lambda_{5},\\
\lambda_{P 1},\lambda_{P 2}
\end{array}
\Longleftrightarrow
\begin{array}{c}v, m_{h},m_{A},m_{H},m_{H^{\pm}}, m_{a},m_{\chi},\\
\tan\beta,\cos (\beta-\alpha),\sin\theta,\\
y_{\chi},\lambda_{3},\lambda_{P 1},\lambda_{P 2}
\end{array}
\label{eq:9}
\end{equation}

When $\sin\theta \approx 0$, the additional pseudoscalar messenger $a$ is mainly composed of $P$. The parameters appearing on the right side of Eq.\eqref{eq:9} are used as the inputs for the analysis of the 2HDM + a model. Generally, the alignment limit ($\cos(\beta - \alpha) = 0$) is assumed. Then the state $h$ corresponds to the 125 GeV Higgs boson, and $H$ represents the CP-even Higgs boson of the BSM. In this paper, we take the same assumptions as those taken by experimental studies~\cite{ATLAS:2020yzc,ATLAS:2022znu, ATLAS:2023rvb} for evading constraints and simplicity. The coupling of $a\chi\bar{\chi}$ and quartic couplings are assumed as $y_{\chi}=1$ and $\lambda_3=\lambda_{P 1}=\lambda_{P 2}=3$. And the mass relation of heavy Higgs bosons are taken as $m_{A}=m_{H}>m_{H^{\pm}}$. Integrating out the heavy $A$ and $H$, there are five free parameters, including mass parameters $m_a$, $m_{\chi}$, $m_{H^{+}}$, mixing parameter $\sin\theta$ and the VEV ratio $\tan\beta$.

The Yukawa couplings in the 2HDM+a model are given by
\begin{eqnarray}
- {\cal L}_Y &=& \frac{m_f}{v}~ y_h^f~h\bar{f}f+\frac{m_f}{v}~y_H^f~H\bar{f}f\nonumber\\
&&-i\frac{m_u}{v}\kappa_u c_\theta~A \bar{u} \gamma_5 u + i\frac{m_d}{v}\kappa_d c_\theta~A \bar{d} \gamma_5 d+ i\frac{m_\ell}{v}\kappa_\ell c_\theta~ A \bar{\ell} \gamma_5 \ell\nonumber\\
&&-i\frac{m_u}{v}\kappa_u s_\theta~a \bar{u} \gamma_5 u + i\frac{m_d}{v}\kappa_d s_\theta~a \bar{d} \gamma_5 d+ i\frac{m_\ell}{v}\kappa_\ell s_\theta~ a \bar{\ell} \gamma_5 \ell\nonumber\\
&&+ H^+~ \bar{u} ~V_{\rm CKM}~ (\frac{\sqrt{2}m_d}{v}\kappa_d P_R  - \frac{\sqrt{2}m_u}{v}\kappa_u P_L)  d+ h.c.\nonumber\\
& &+ \frac{\sqrt{2}m_\ell}{v}\kappa_\ell H^+~\bar{\nu} P_R  e + h.c.~,\label{yuka-coupling}
\end{eqnarray}
where $y_h^f=\sin(\beta-\alpha)+\cos(\beta-\alpha)\kappa_f$ and $y_H^f=\cos(\beta-\alpha)-\sin(\beta-\alpha)\kappa_f$. In the type II model, the Yukawa factors are $\kappa_u=1/\tb$ and $\kappa_d=\kappa_{\ell}=-\tb$, where as in the type I case one has $\kappa_f=1/\tb$ for all fermions $f=u,d,\ell$. In below calculation, we focus on the Type I 2HDM+a, which constraints are less stringent.

\begin{figure}[htbp]
  \centering
  \begin{minipage}[b]{0.48\textwidth}
    \centering
    \includegraphics[width=\textwidth]{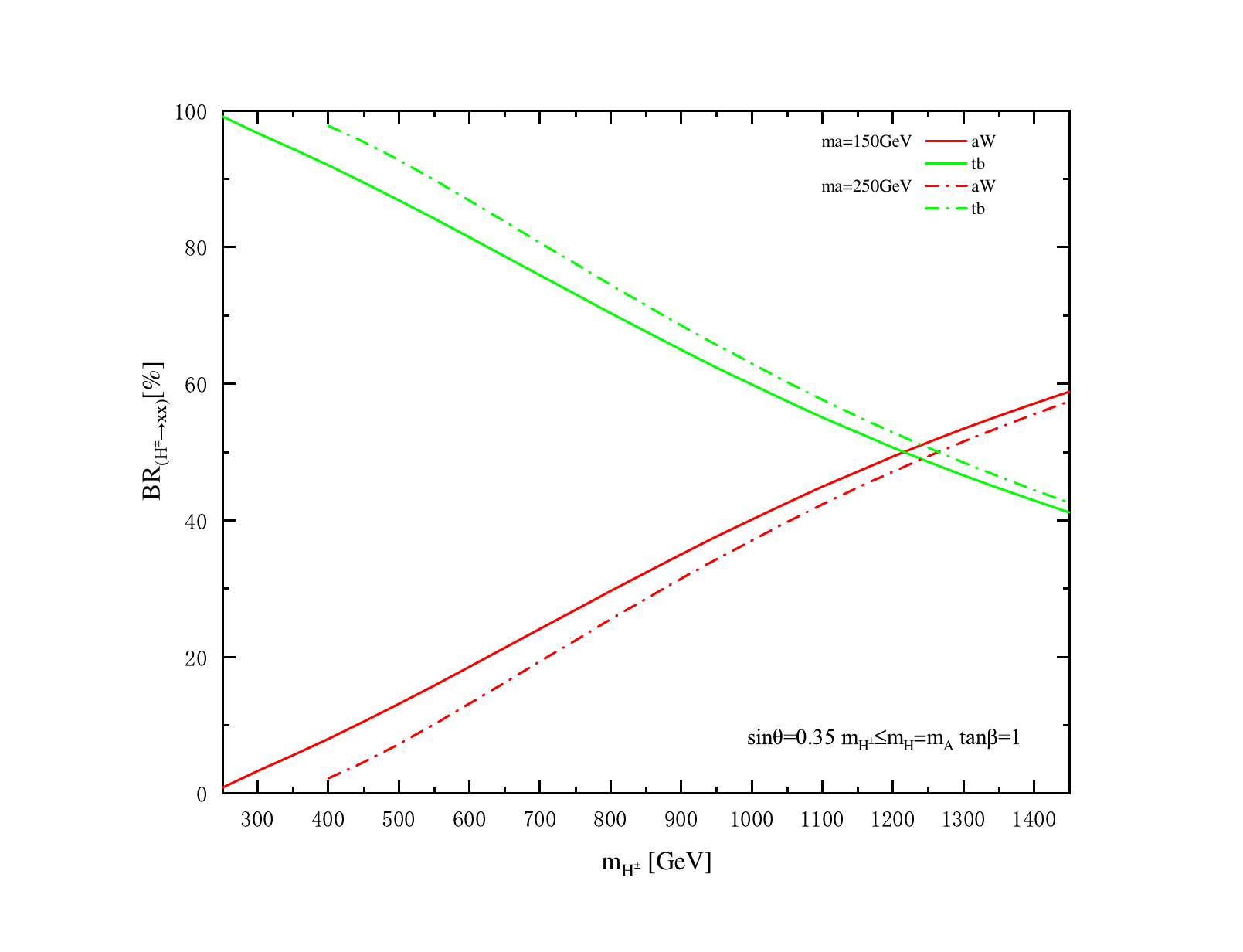}
  \end{minipage}
  \hfill
  \begin{minipage}[b]{0.48\textwidth}
    \centering
    \includegraphics[width=\textwidth]{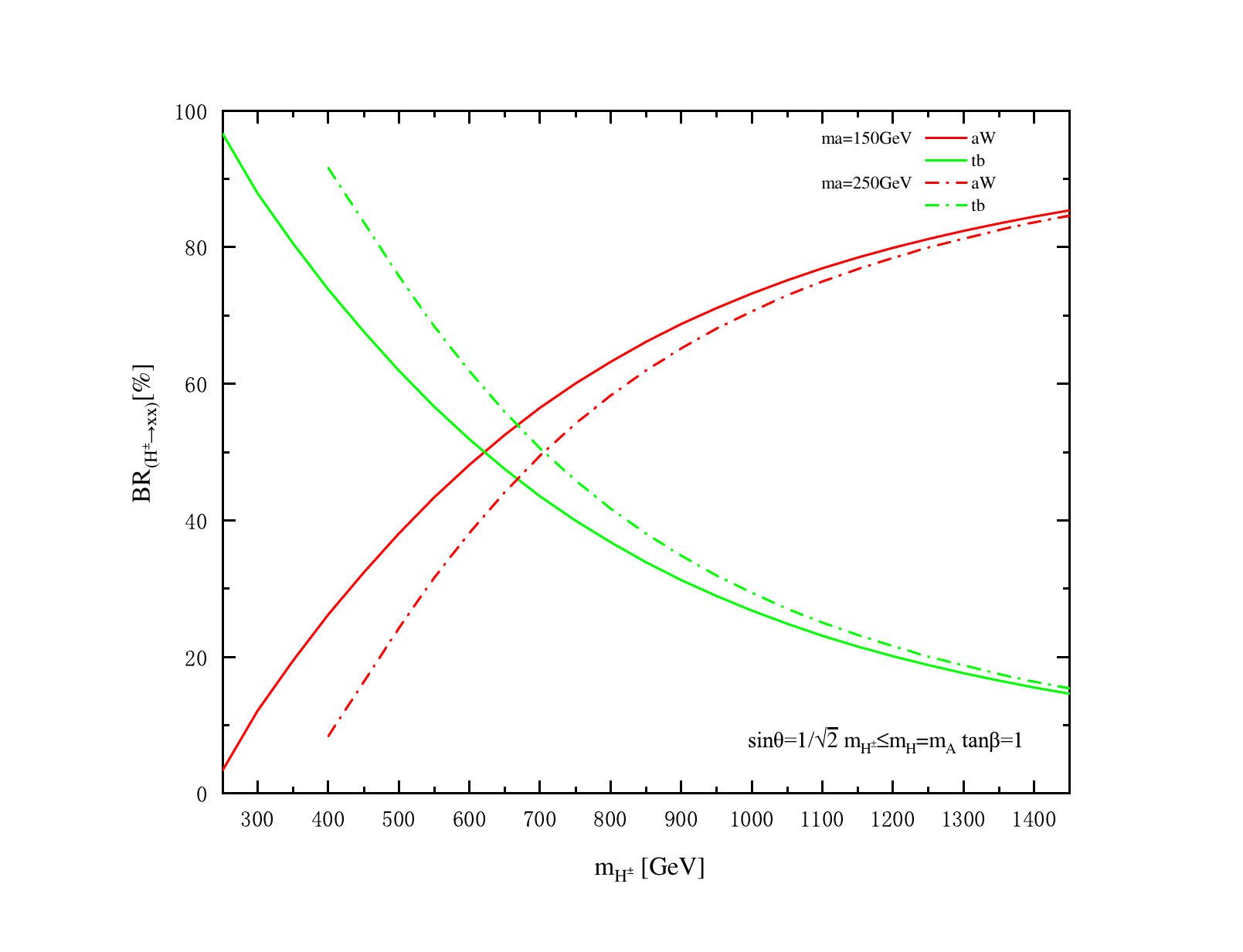}
  \end{minipage}
 \caption{\label{fig:BR} Charged Higgs branching ratios as a function of its mass $m_{H^{\pm}}$ with different values of $m_a$ (solid line for $m_a$=150, dashed line for $m_a=250$). Here, the red curves represent the branching ratio of the $H^{\pm}\to aW$, while the green curves are relevant to $H^{\pm}\to tb$ channel. Two typical value of $\sin\theta=0.35$ and $\sin\theta=1/\sqrt{2}$ are shown in left and right, respectively. In both subfigures, $m_A=m_H=1.5\text{TeV}$ and $\tan\beta=1$ are fixed.}
\end{figure}

\section{Analysis of \texorpdfstring{$e^+e^- \to H^+ H^- \to tb a W \to 4j + 2b +  E_T^{\text{miss}}$}{e+e- → H+H- → tb a W → 4j + 2b + ET miss} at CLIC}
\subsection{Charged Higgs boson consideration}

In the framework of the 2HDM+a, the introduction of pseudoscalar singlet $a$ induces rich Higgs phenomena. One special new decay mode of charged Higgs boson decay channel $H^+ \to aW$ is opened, which is a mode unique to this model.

\vspace{-0.2cm}
\hspace{-1.2cm}
\begin{figure}[hbp]
\centering
  \begin{minipage}{0.46\textwidth}
  \centering
\includegraphics[width=\textwidth]{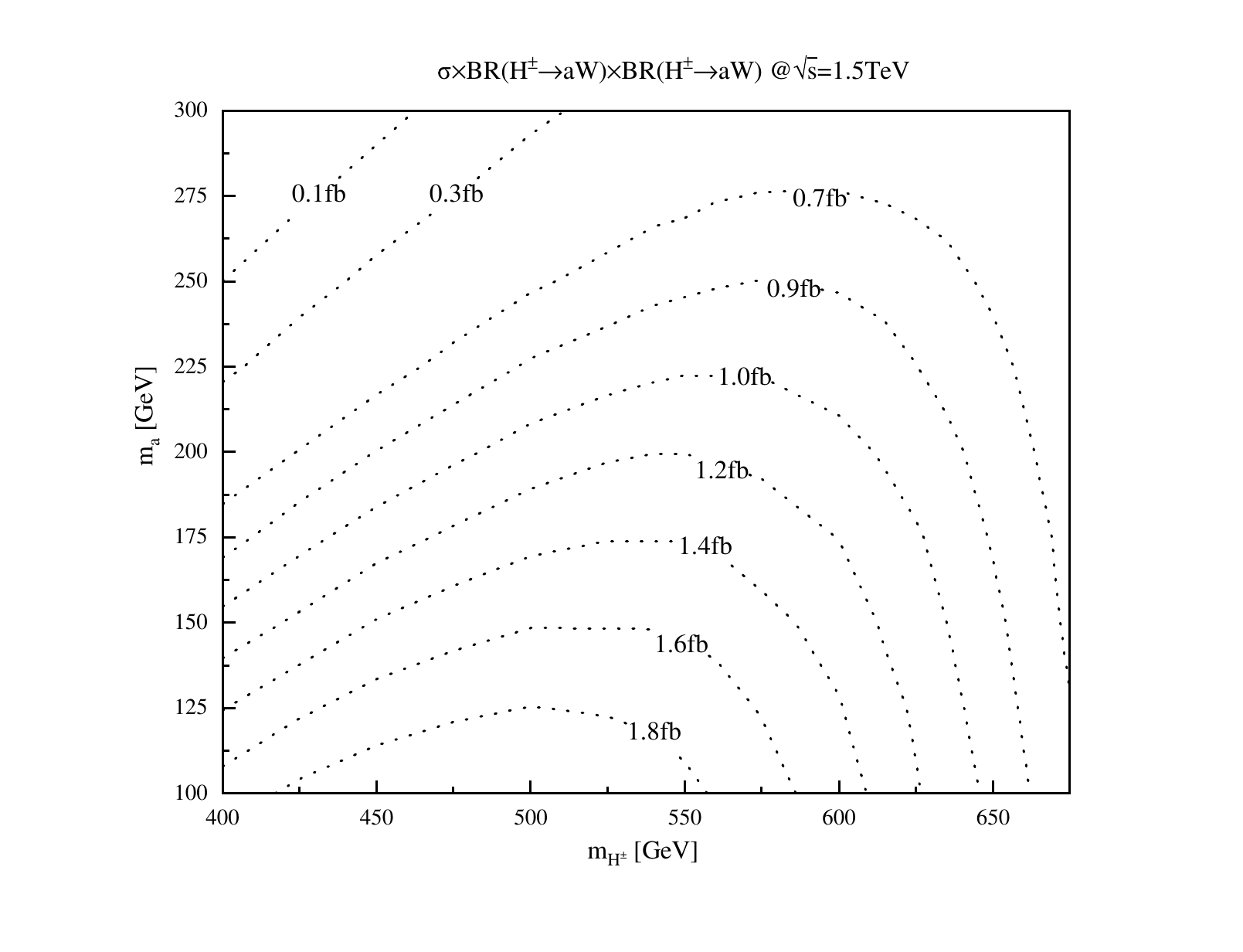}
\end{minipage}
   \hspace{0.1cm}
  \begin{minipage}{0.46\textwidth}
    \centering
    \includegraphics[width=\textwidth]{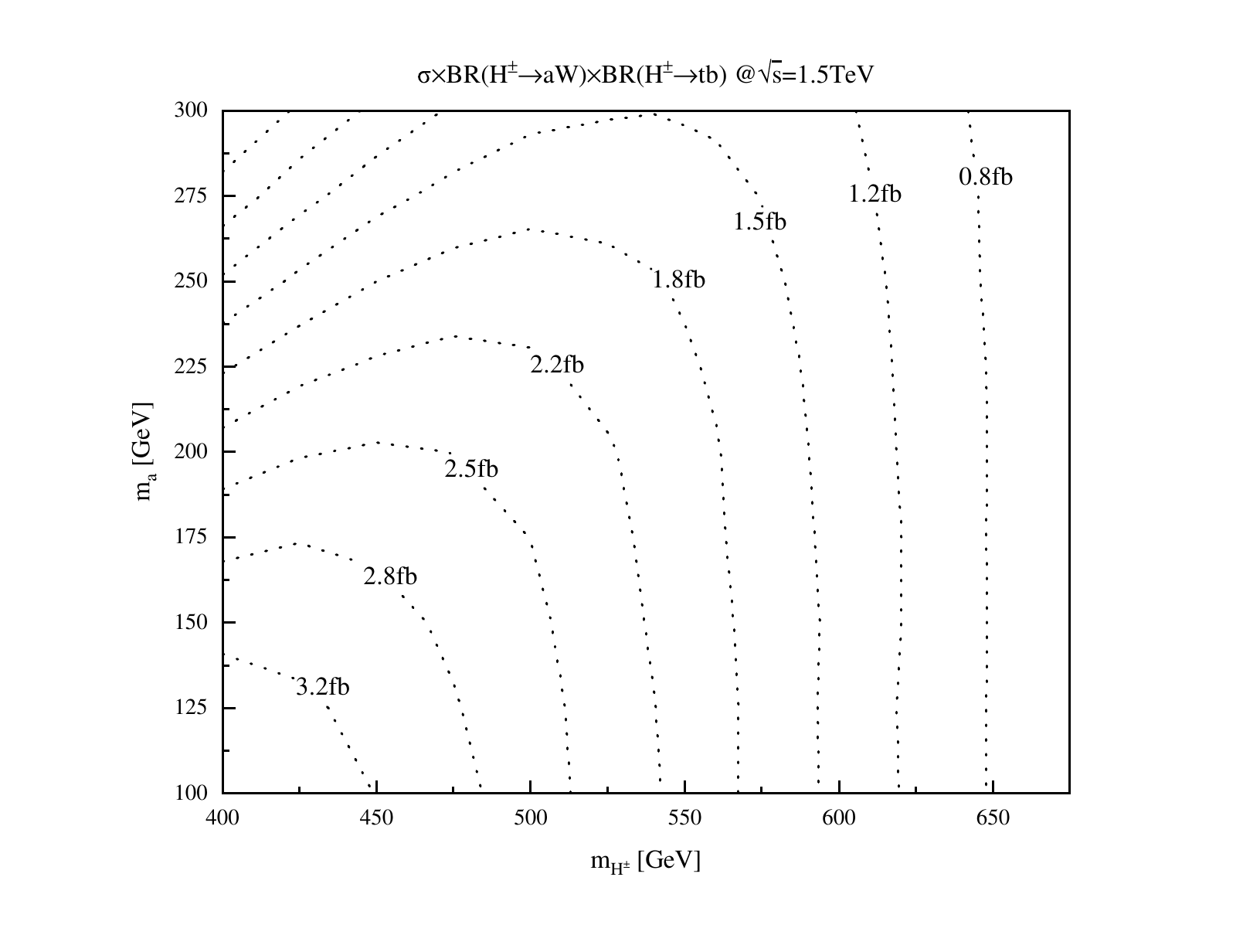}
  \end{minipage} 
  \vspace{0cm}  
  \begin{minipage}{0.46\textwidth}
    \centering
    \includegraphics[width=\textwidth]{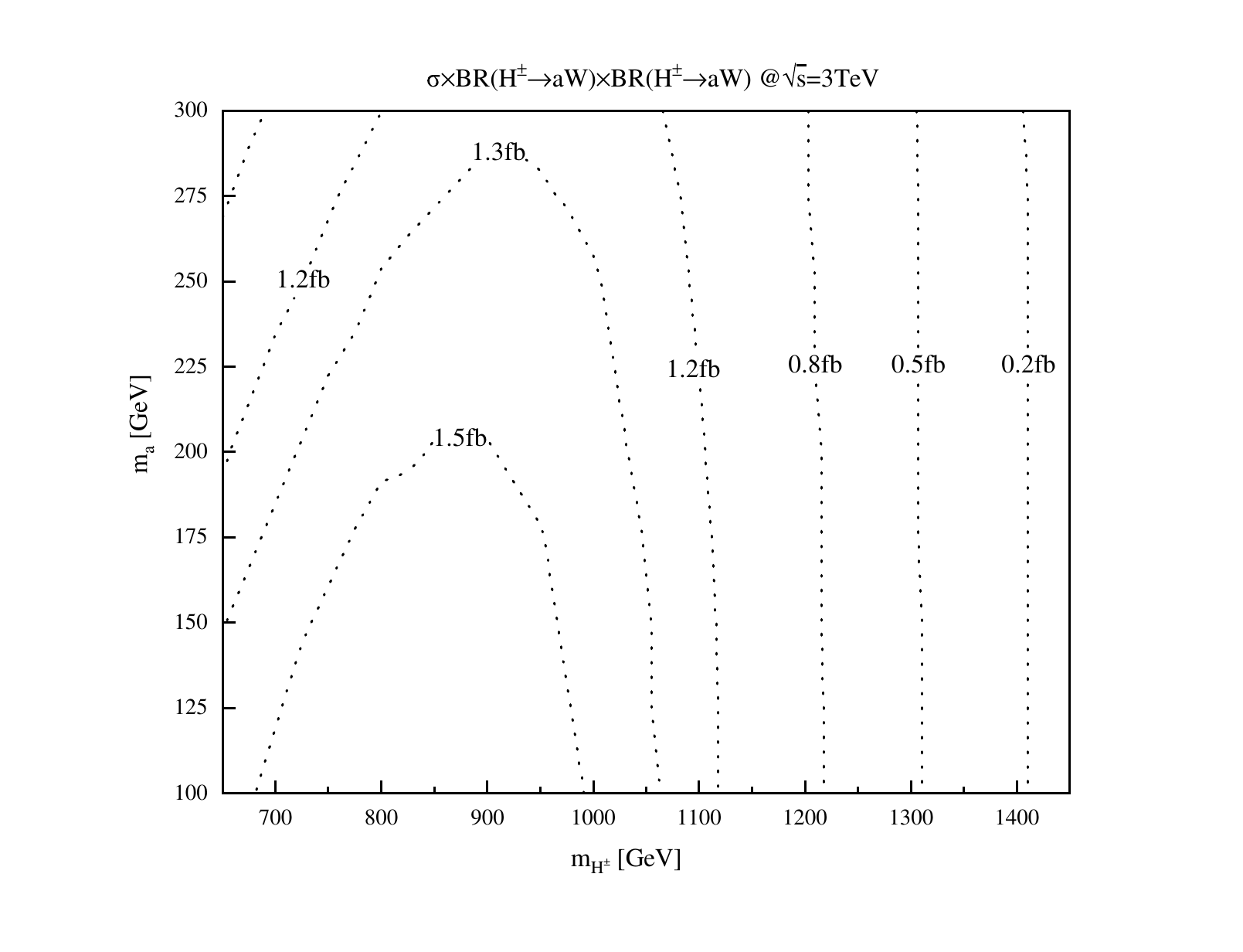}
  \end{minipage}
  \hspace{0.1cm}
  \begin{minipage}{0.46\textwidth}
    \centering
    \includegraphics[width=\textwidth]{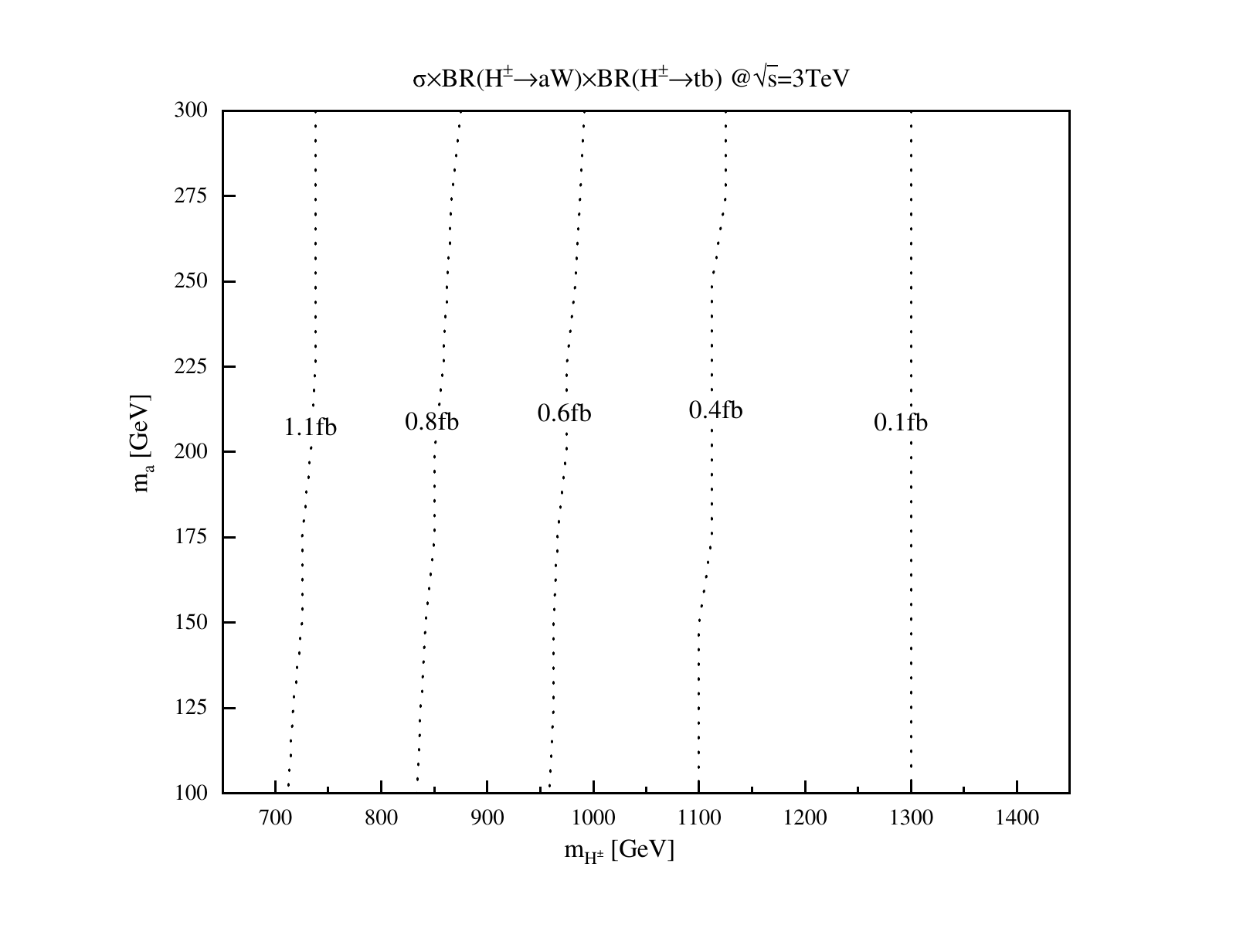}
  \end{minipage}
  \caption{\label{fig:2}The cross section times branching ratio of $e^+e^- \to H^+H^- \to aW^+aW^-$/$t\bar{b}aW^-$ as a function in $(m_{H^\pm}, m_a)$ plane in CLIC. Other parameters are set as: $m_A = m_H = 1.5\,\text{TeV}$, $y_\chi = 1$, $\lambda_1 = \lambda_2 = \lambda_3 = 3$, $\tan\beta = 1$, $\sin\theta = 1/\sqrt{2}$, $\sin(\beta - \alpha) = 1$, $m_\chi = 100\,\text{GeV}$.}
  \label{fig:XtimesBR}
\end{figure}

Considering in the alignment limit,
the partial decay widths of the charged Higgs in 2HDM+a are given by the following expressions~\cite{Bauer:2017ota, Argyropoulos:2024yxo}:

\begin{align}
\Gamma (H^+ \rightarrow t\bar{b}) &= \frac{N_c^t |V_{tb}|^2 \cot ^2 \beta}{8 \pi} \frac{m_{{t}}^2}{v^2} m_{H^\pm} \left( 1 - \frac{m_{{t}}^2}{m_{H^\pm}^2} \right)^2, \\
\Gamma (H^+ \rightarrow HW^+) &= \frac{1}{16 \pi} \frac{\lambda^{3/2}(m_{H^\pm}, m_H, m_W)}{m_{H^\pm}^3 v^2}, \\
\Gamma (H^+ \rightarrow AW^+) &= \frac{1}{16 \pi} \frac{\lambda^{3/2}(m_{H^\pm}, m_A, m_W)}{m_{H^\pm}^3 v^2} \cos ^2 \theta, \\
\Gamma (H^+ \rightarrow aW^+) &= \frac{1}{16 \pi} \frac{\lambda^{3/2}(m_{H^\pm}, m_a, m_W)}{m_{H^\pm}^3 v^2} \sin ^2 \theta
\end{align}

with\[
\lambda(m_1, m_2, m_3) = \left( m_1^2 - m_2^2 - m_3^2 \right)^2 - 4m_2^2m_3^2
\]

In the alignment limit, the $H^+ \rightarrow hW^+$ decay mode does not exist. With an assumption of $m_A = m_H \geq m_{H^\pm}$, the $H^+ \to H W^+$ / $A W^+$ decays are kinematically forbidden and only two decay modes $H^+\rightarrow t\bar{b}$ and $H^+\rightarrow aW^+$ are left. And for $H^+ \rightarrow t\bar{b}$, terms of order $\mathcal{O}\left(\frac{m_b^2}{m_{H^\pm}^2}\right)$ are neglected in the decay width expression. It is remarked that these expressions are the same for Type II 2HDM+a and Type I 2HDM+a in the limit of alignment and small $\tan\beta$.
In this paper, we choose the type-I 2HDM+a model as a show case to conduct a numerical analysis in the alignment limit and small $\tan \beta$ region. However, the conclusion is also applicable to type-II 2HDM+a.

While both the partial widths grow with increasing charged Higgs mass, the growth of $\Gamma(H^+\rightarrow aW^+)$ is more rapid. As shown in Fig.\ref{fig:BR}, the $BR(H^+ \to aW^+)$ increases with the mass $m_{H^{+}}$ whereas the $BR(H^+ \to t\bar{b})$ exhibit opposite trend. 
Setting $tan\beta=1$ and $m_A=m_H=1.5\text{TeV}$, results for different $sin\theta$ and $m_a$ are presented. For a large value of $\sin\theta=1/\sqrt{2}$, the Br($H^+\rightarrow aW^+$) is much larger and it can reach $74\%$ ($m_a=150\text{GeV}$) or higher for $m_{H^+} \geq 1$TeV.

 In Fig.\ref{fig:XtimesBR}, we present the cross section times relevant branching ratios of $H^+ \to aW^+$ and $H^+ \to t\bar{b}$ in the parameter space of $(m_{H^\pm}, m_a)$ at CLIC with $\sqrt{s}=1.5\text{TeV}$ and with $\sqrt{s}=3 \text{TeV}$. As the increase in the mass of the charged Higgs boson, the values of $\sigma \times \text{BR}$ for the corresponding decay channels exhibit a decreasing trend. Furthermore, for the same value of $m_a$, the phenomenon where the values of $\sigma \times \text{BR}$ are equal may arise at different parameter points of $m_{H^\pm}$. This is due to the fact that when $m_{H^\pm}$ is relatively small, the branching ratio of the decay channel $H^\pm \to aW^\pm$ is low, whereas the production cross-section of charged Higgs pairs is large. Conversely, when $m_{H^\pm}$ is large, the branching ratio of $H^+ \to aW^+$ increases, but the production cross-section of charged Higgs pairs decreases. As a result, the situation occurs where $\sigma \times \text{BR}$ has the same value for different values of $m_{H^\pm}$.

In this paper, we focus on the process $e^+e^- \to H^+H^- \to tb aW$ at CLIC with $\sqrt{s}=1.5\text{TeV}$. This is partially because of the complementarity of BRs for the two decay channels $H^+ \to aW^+$ and $H^+ \to t\bar{b}$ as shown in Fig.\ref{fig:BR}. In addition, the $e^+e^- \to H^+H^- \to aW aW$ channel have less information to carry out reconstruction due to the invisible decay of the mediator $a\to \chi \chi$.  

\subsection{Signal and Background}
In this study, the pair production of charged Higgs followed with the decays of $H^{\pm} \to tb$ and $H^{\mp} \to aW$ at the 1.5 TeV CLIC is explored. Considering following hadronic decay of $t$ and $W$, along with the decay mode of $a\to \chi\chi$, the complete signal channel under investigation is 
\begin{equation*}
e^+ e^- \to H^+ H^- \to t baW \to 4j+2b+E_T^{miss}.
\end{equation*}

Consequently, the final state features of this signal channel consist of 4 jets, 2 b-jets, and missing transverse energy $E_T^{\text{miss}}$. In this channel, the large branching ratio of hadronic decay can compensate for the relatively small production rate. The intricate collision environment at the LHC, dominated by substantial QCD backgrounds, poses significant challenges for signal reconstruction in high-jet-multiplicity events with low missing energy‌. However, the clean collision environment and excellent calorimeter performance make CLIC a powerful tool to exploring this channel.

In this research, the signal and background events are initially in parton level generated with MadGraph5\_aMC@NLO. And then parton showering and hadronization are implemented via Pythia8. Detector simulations are conducted with Delphes. The Valencia jet algorithm (VLC) designed for future lepton colliders is employed with jet radius parameter $R = 0.5$. Finally, a cut-based analysis is performed using MadAnalysis 5.

For simplicity, the DM mass is fixed as $m_\chi = 10 \text{GeV}$ and four benchmark points are taken to illustrate the prospects for detecting as follows:
\begin{itemize}
    \item BP1: $m_{H^+} = 400 \text{GeV}$, $m_a = 250 \text{GeV}$, $\tan\beta = 1$, $\sin\theta = 1/2$  
    \item  BP2: $m_{H^+} = 400 \text{GeV}$, $m_a = 250 \text{GeV}$, $\tan\beta = 1$, $\sin\theta = 1/\sqrt{2}$ 
    \item BP3: $m_{H^+} = 600 \text{GeV}$, $m_a = 350 \text{GeV}$, $\tan\beta = 3$, $\sin\theta = 1/2$  
    \item  BP4: $m_{H^+} = 600 \text{GeV}$, $m_a = 350 \text{GeV}$, $\tan\beta = 3$, $\sin\theta = 1/\sqrt{2}$ 
\end{itemize}

The dominant background processes encompass the production processes of $t\bar{t}h$, $t\bar{t}Z$, $t\bar{t}j$, and $t\bar{t}$.
\begin{itemize}
    \item $e^+e^- \to t\bar{t}$ with $t\bar{t} \to W^\pm b$ and $W^\pm \to jj$
    \item $e^+e^- \to t\bar{t}j$ with $t\bar{t} \to W^\pm b$ and $W^\pm \to jj$
    \item $e^+e^- \to t\bar{t}h$ with $t\bar{t} \to W^\pm b$, $W^\pm \to jj$, and $h \to b\bar{b}$
    \item $e^+e^- \to t\bar{t}Z$ with $t\bar{t} \to W^\pm b$, $W^\pm \to jj$, and $Z \to jj$
\end{itemize}

\begin{figure}[ht]
    \centering
    \begin{minipage}[b]{0.48\textwidth}
        \centering
        \includegraphics[width=\textwidth]{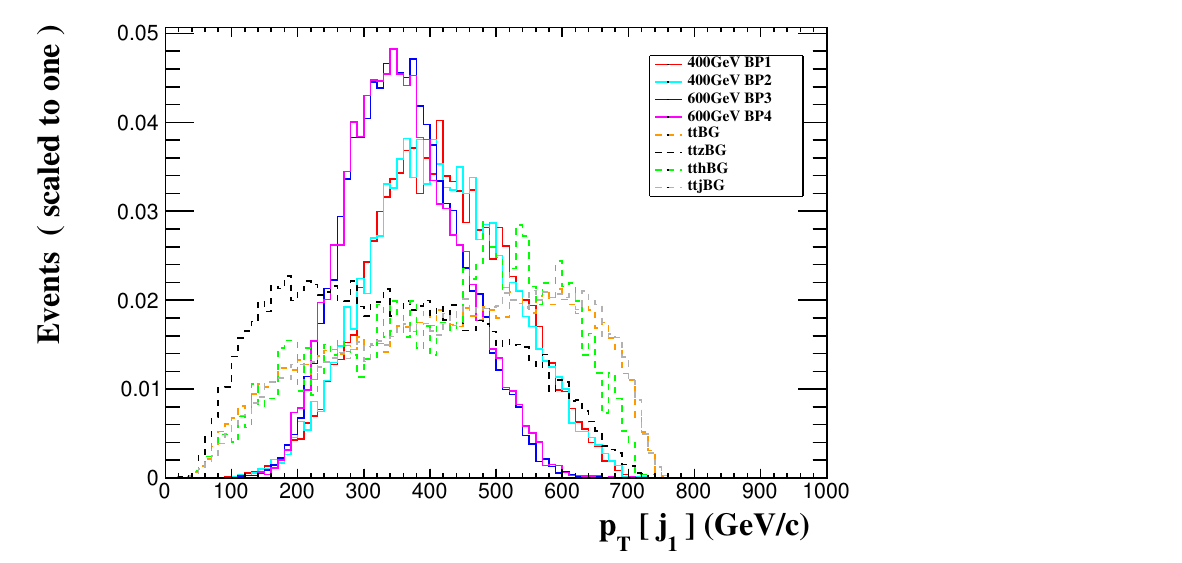}
    \end{minipage}
    \hfill
    \begin{minipage}[b]{0.48\textwidth}
        \centering
        \includegraphics[width=\textwidth]{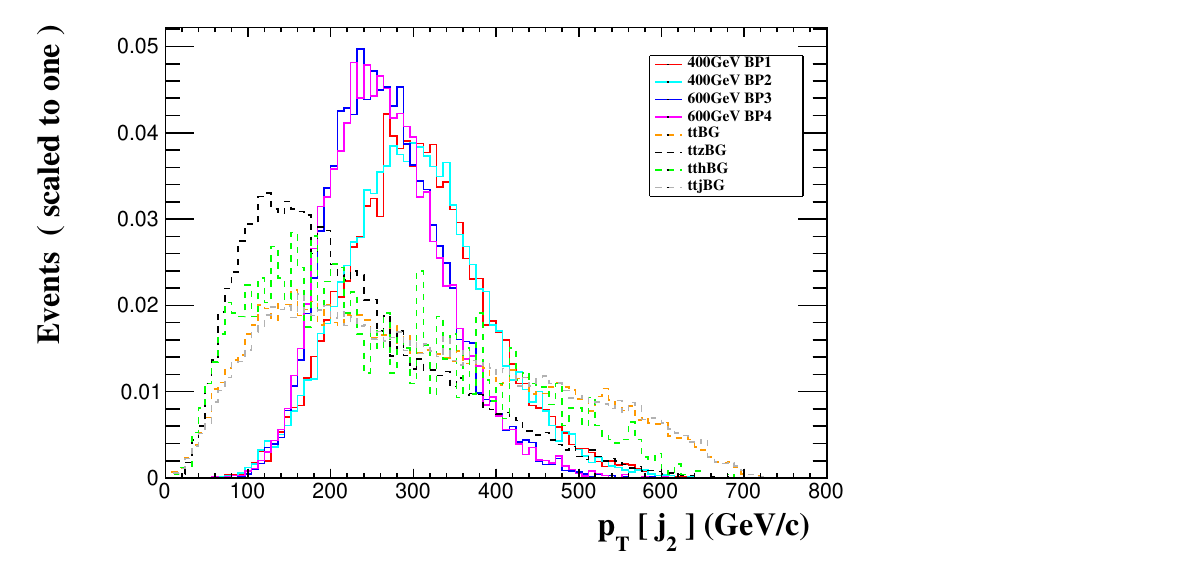}
    \end{minipage}
    \vspace{0.5cm}
    \begin{minipage}[b]{0.48\textwidth}
        \centering
        \includegraphics[width=\textwidth]{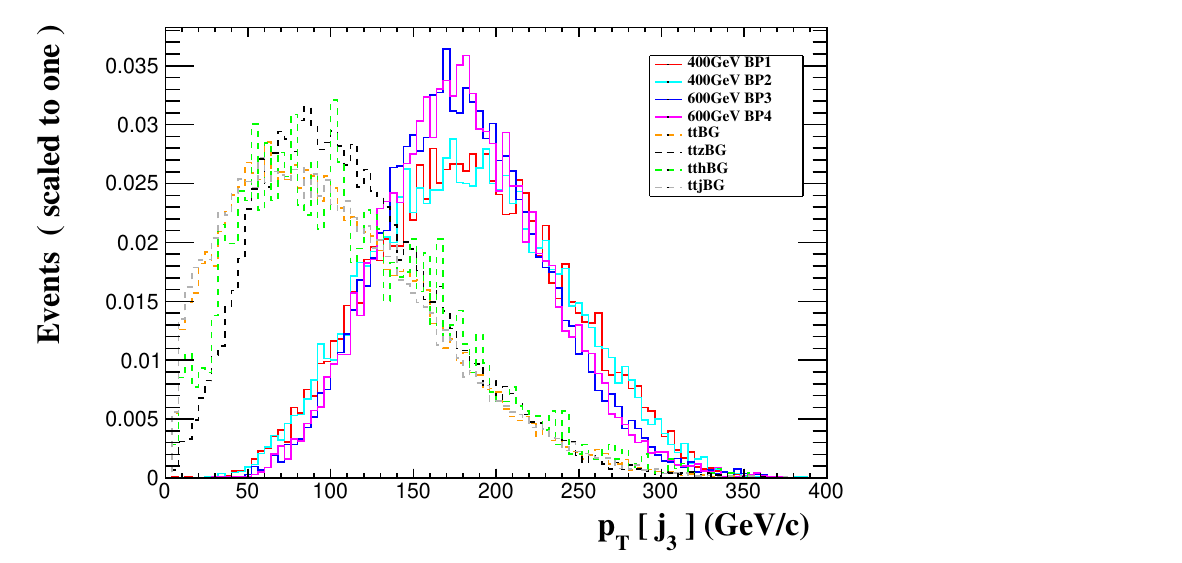}
    \end{minipage}
    \hfill
    \begin{minipage}[b]{0.48\textwidth}
        \centering
        \includegraphics[width=\textwidth]{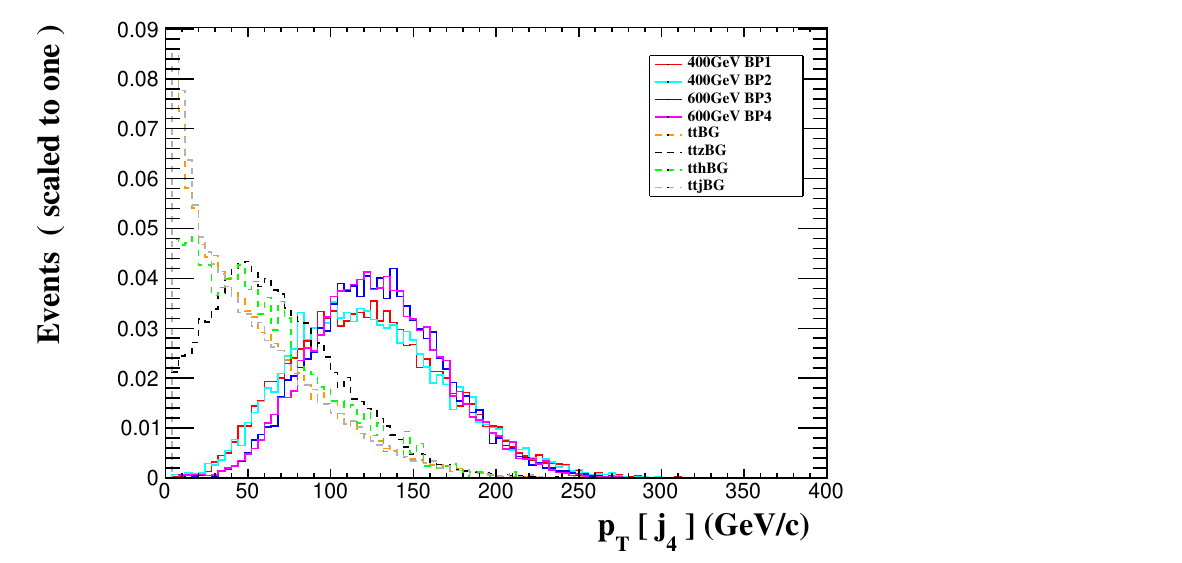}
    \end{minipage}
    \caption{The normalized distributions for jet ($p_T$) for the signal and backgrounds. Four benchmark points are considered and labeled as BPs.
    }
    \label{fig:ptj}
\end{figure}

To simulate detector acceptance and provide effective trigger, we choose the basic cuts at parton level for the signals and SM backgrounds as follows:
\begin{equation*}
p_{\text{T}}(j) > 5\,\text{GeV} \quad \text{and} \quad |\eta(j)| < 5,\quad  E_{\text{T}}^{\text{miss}} > 10\,\text{GeV}
\end{equation*}
where $p_{\text{T}}(j)$ and $\eta(j)$, $E_{\text{T}}^{\text{miss}}$are the transverse momentum of jets, pseudo-rapidity of jets and transverse missing energy, respectively. After selecting events with a final state containing more than 4 jets, an additional 2 b-jets, and no leptons, we sort these final jets by $p_T$ order and show the normalized kinematic distributions of signals and backgrounds in Fig.\ref{fig:ptj}.
For signal events, the jets are inclined to have a large $P_T$ than those in backgrounds. 

In order to extract the signal from the backgrounds, a set of improved cuts are adopted according to signal's features. These cuts are taken as follows: 

\begin{itemize}
    \item \textbf{Cut-1:} The scalar sum of all transverse momentum of all jets is required to be larger than 1000 GeV and two tagged $b$-jets are required with $b$-tagging efficiency $\epsilon_b = 80\%$:
    \[N(\text{$b$-jet}) = 2\quad\text{and}\quad H_{\text{T}} > 1000\,\text{GeV}.\]
    \item \textbf{Cut-2:} Furthermore, the jets are combined to form candidate top and W bosons by minimizing a defined chi-squared criterion and $\chi_{min}<15$ is taken. The $\chi^2$ is defined as
     \[
    \chi^2 =  \frac{(m_{ijr} - m_{t})^2}{\sigma_{t}^2} + \frac{(m_{ij} - m_{W})^2}{\sigma_W^2} + \frac{(m_{kl} - m_{W})^2}{\sigma_W^2}.
    \]
    Here, $m_{ij}$ and $m_{kl}$ denote the invariant mass of the jet pairs used to reconstruct the $W$ candidates, $m_{ijr}$ is the invariant mass of two jets and one $b$-jet. And $\sigma_{t}=15$ GeV and $\sigma_{W}=10$ GeV are the estimated invariant mass resolutions for top and $W$ candidates, respectively. 
    \item \textbf{Cut-3:} Transverse momentum of the reconstructed top candidates are required as: \[p_{\text{T}}(t) < 400\,\text{GeV}.\]
    \item \textbf{Cut-4:} A mass window cut for reconstructed $H^+$ from $t b$ is employed:
    \[
    m_{H^+}^{\text{rec} }= m_{H^+} \pm 60\,\text{GeV}
    \]
\end{itemize}

\begin{figure}
    \centering
    \includegraphics[width=0.6\linewidth]{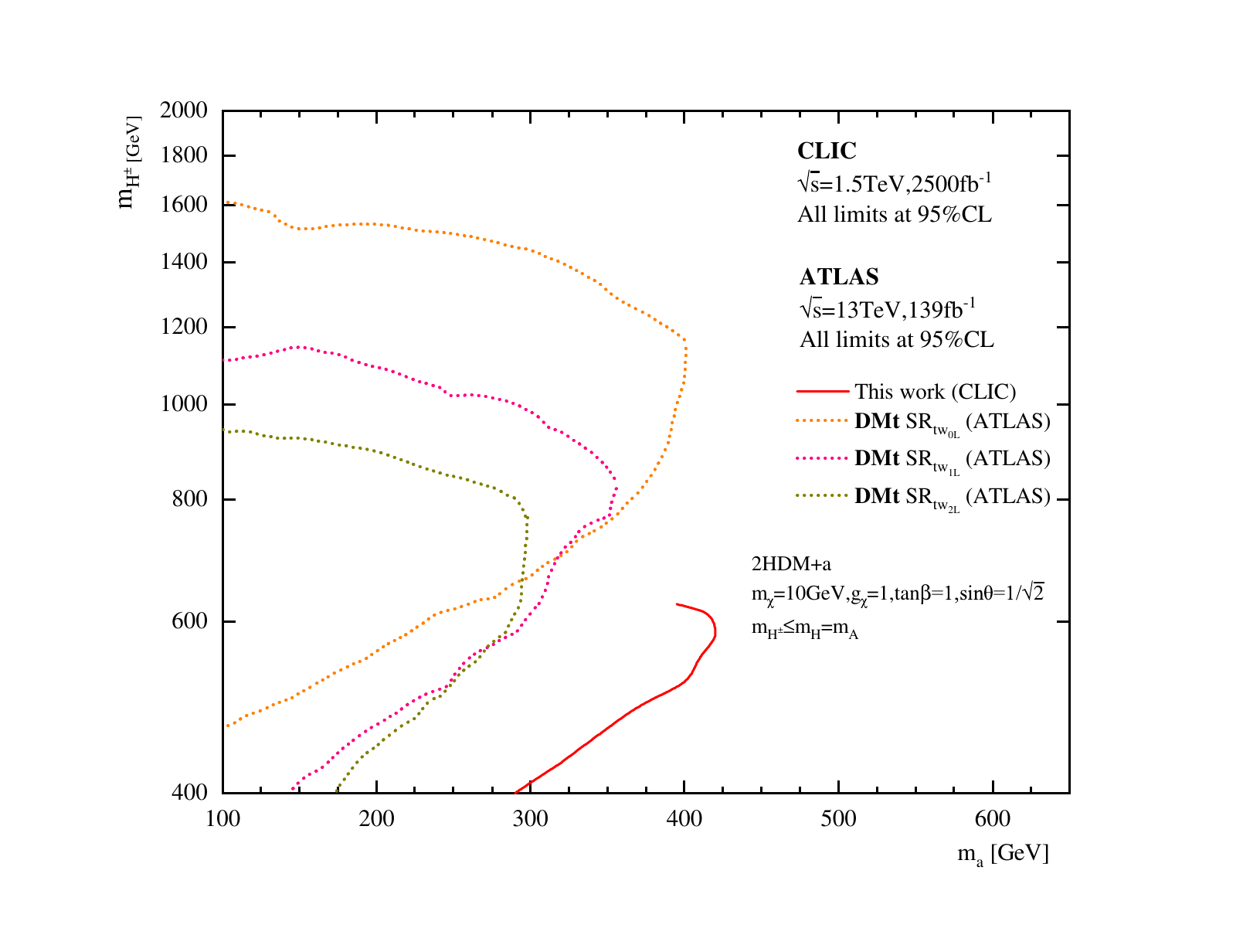}
    \caption{The observed exclusion contours as a function of $(m_a, m_{H^\pm})$ plane. The figure shows a comparison between the ATLAS results and the results of our work. The red solid line in the figure is the result of our work.}
    \label{fig:exlimits}
\end{figure}

The comparison of the cut efficiencies between the signal and backgrounds is summarized in \textbf{Table~\ref{tab:1}}. Four benchmark points are taken to show. It is found that all the SM backgrounds are efficiently suppressed after imposing the cuts, while the signal events are kept in a relatively large efficiency. Taking an integrated luminosity of 2500 fb$^{-1}$, a  statistical significance $S/\sqrt{S+B}=6.18(6.95) $ for BP1 (BP3) can be achieved. Furthermore, taking $sin\theta = 1/\sqrt{2}$ and $tan \beta=1$, we analyze statistical significances for this signal channel with variable $m_{H^{+}}$ and $m_{a}$. The results are shown in the Fig.~\ref{fig:exlimits}. For the charged Higgs with a mass smaller than 600 GeV, the statistical significance can reach several $\sigma$ in most of the parameter plane. For the heavier charged Higgs, it is difficult to detect the signal due to the limited production rates. 

\begin{table}[hb]
    \centering
    \captionsetup{skip=10pt} 
    \caption{The cut flow for the signal and backgrounds where the cut efficiencies for four benchmark points are shown. The statistical significances are calculated for an integrated luminosity of $2500\,\text{fb}^{-1}$. The values in the brackets represent the cross section after Cut-4 with $m_{H^+}^{\text{rec} }= 600 \pm 60\,\text{GeV}$.}
    \label{tab:1}
    \renewcommand{\arraystretch}{1.3} 
   
    \tabcolsep=4pt
    \begin{tabular}{l|cc|cc|cccc}
      \hline
        & \multicolumn{4}{c|}{Signals} & \multicolumn{4}{c}{Backgrounds} \\
         \cline{2-9}
        Cuts & \multicolumn{2}{c|}{400\,\text{GeV}} & \multicolumn{2}{c|}{600\,\text{GeV}} \\
        & BP1 & BP2 & BP3 & BP4 & $\bar{t}t h$ & $\bar{t}t Z$ & $\bar{t}t j$ & $\bar{t}t$ \\
        \hline
        Basic & 0.566 & 0.698 & 0.838 & 1.522 & 0.370 & 1.053 & 11.924 & 13.323 \\
        Cut-1 & 0.266 & 0.334 & 0.376 & 0.690 & 0.010 & 0.165 & 2.704 & 2.933 \\
        Cut-2 & 0.122 & 0.154 & 0.220 & 0.416 & 0.004 & 0.093 & 1.125 & 1.204 \\
        Cut-3 & 0.106 & 0.134 & 0.198 & 0.376 & 0.003 & 0.060 & 0.516 & 0.587 \\
        Cut-4 & 0.058 & 0.074 & 0.068 & 0.134 & 0.0005(0.0003) & 0.011(0.008) & 0.078(0.080) & 0.073(0.083) \\
        \hline
        $S/\sqrt{S+B}$ & 6.18 & 7.61 & 6.95 & 12.13 &  &  &  &  \\
        \hline
    \end{tabular}
\end{table}

\begin{figure}[hb]
    \centering
    \includegraphics[width=0.6\linewidth]{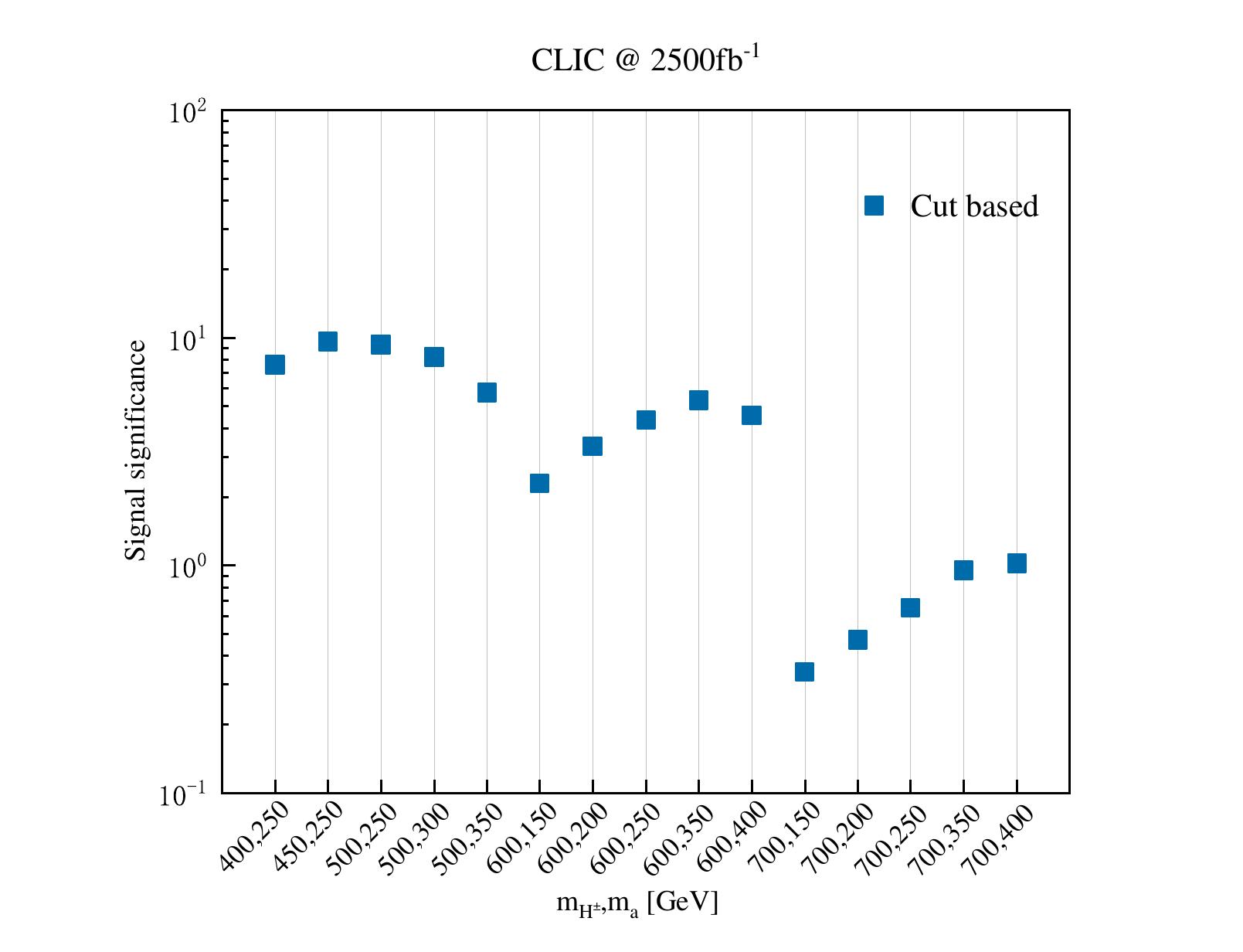}
    \caption{The signal significances obtained from the cut based analysis for a set of points with an integrated luminosity of 2500 fb$^{-1}$}
    \label{fig:significances}
\end{figure}

\section{Conclusions}

As a natural framework that accommodates a DM candidate while addressing some open issues of the SM, 2HDM+a induce rich collider phenomena and is recognized by the LHC dark Matter Working Group as a benchmark model. In this study, we explore the pair production of charged Higgs boson via the process $e^-e^+ \to H^+H^- \to tb aW\to 4j+2b+E_T^{miss}$ at the 1.5 TeV CLIC. 

The detailed simulation is performed for the signal and the main SM backgrounds. The cut based analysis find that the signal significance of charged Higgs can reach $5\sigma$ in certain parameter space.  
In Fig.~\ref{fig:exlimits}, we also present the exclusion limit at 95\% confidence level at 1.5 TeV CLIC with an luminosity of $2500\,\text{fb}^{-1}$, comparing it with the current exclusion curves obtained by ATLAS group based on 13 TeV collision dataset with $139\,\text{fb}^{-1}$ luminosity~\cite{ATLAS:2022znu}. Our calculation of the exclusion limit is complementary to the LHC existing bounds and provide valuable insights for future CLIC experiment. It is evident that CLIC exhibits higher sensitivity to the relatively small $m_{H^{\pm}}-m_{a}$ region for $m_{H^{\pm}}$ in the range of 400 GeV $\sim$ 650 GeV. The intricate collision environment at the LHC presents significant challenges for signal reconstruction in high-jet-multiplicity events with low missing energy‌, whereas CLIC demonstrates superior capability in capturing signal events within this parameter space. 
However, for larger charged Higgs mass, the limited center-of-mass energy of 1.5 TeV is not possible to produce signal events. For the center-of-mass energy of 3 TeV, the production cross section for charged Higgs boson is tiny due to the s-channel suppression.

\section*{Acknowledgements}

This work was supported in part by the National Natural Science Foundation of China under Grants No. 12575106, No. 12147214, No. 11905093, and No. 11975013, and the Basic Research Project of Liaoning Provincial Department of Education for Universities under Grants No. LJKMZ20221431 and Teaching Reform Research Project for graduates of Liaoning Normal University, and by the Projects No. ZR2024MA001 and
No. ZR2023MA038 supported by Shandong Provincial 
Natural Science Foundation.

\bibliography{sample}
\bibliographystyle{CitationStyle}
\end{document}